\documentclass[11pt]{article}

\newlength{\vshift}
\newlength{\hshift}
\usepackage{amssymb,amsopn}

\def\nn{\nonumber }

\def\be{\beta}
\def\a{\alpha}

\def\g{\gamma}

\def\ds{\stackrel{\star}{,}}

\def\tr{{\rm Tr}}

\def\nn{\nonumber}
\def\be{\begin{equation}}             \def\ee{\end{equation}}
\def\ba#1{\begin{array}{#1}}          \def\ea{\end{array}}
\def\bea{\begin{eqnarray} }           \def\eea{\end{eqnarray} }
\def\beann{\begin{eqnarray*} }        \def\eeann{\end{eqnarray*} }
\def\beal{\begin{eqalign}}            \def\eeal{\end{eqalign}}
             
\def\bsubeq{\begin{subequations}}     \def\esubeq{\end{subequations}}
\def\bitem{\begin{itemize}}           \def\eitem{\end{itemize}}
                    
\def\pa{\partial}
\def\a{\alpha}
\def\b{\beta}

\def\d{\delta}
\def\g{\gamma}

\def\k{\kappa}
\def\l{\lambda}
\def\m{\mu}
\def\n{\nu}

\def\r{\rho}                    
\def\s{\sigma}                  

\def\L{\Lambda}

\begin{document}

\begin{titlepage}

$\,$

\vspace{1.5cm}
\begin{center}

{\LARGE{\bf Noncommutative $SO(2,3)$ gauge theory and noncommutative gravity}}

\vspace*{1.3cm}

{{\bf Marija Dimitrijevi\' c and
Voja Radovanovi\' c}}

\vspace*{1cm}

University of Belgrade, Faculty of Physics\\
Studentski trg 12, 11000 Beograd, Serbia \\[1em]

\end{center}

\vspace*{2cm}

\begin{abstract}

In this paper noncommutative gravity 
is constructed as a gauge theory of the noncommutative
$SO(2,3)_\star$ group, while the noncommutativity is canonical (constant). The
Seiberg-Witten map is used to express noncommutative fields in terms of the
corresponding commutative fields. The commutative limit of the model is
the Einstein-Hilbert action with the cosmological constant term and the
topological
Gauss-Bonnet term. We calculate the second order correction to this model and
obtain terms that are of zeroth to fourth power in the curvature
tensor and torsion. Trying to relate our results with $f(R)$ and $f(T)$
models, we analyze different limits of our model. In the limit of big
cosmological constant and vanishing torsion we obtain a $x$-dependent
correction to the cosmological constant, i.e. noncommutativity leads to a
$x$-dependent cosmological constant. We also discuss the limit
of small cosmological constant and vanishing torsion and the teleparallel limit. 

\end{abstract}
\vspace*{1cm}


\vspace*{1cm}
\quad\scriptsize{eMail:
dmarija,rvoja@ipb.ac.rs}
\vfill

\end{titlepage}\vskip.2cm

\newpage
\setcounter{page}{1}
\newcommand{\Section}[1]{\setcounter{equation}{0}\section{#1}}
\renewcommand{\theequation}{\arabic{section}.\arabic{equation}}

\section{Introduction}

\noindent General Relativity (GR) is widely accepted as a classical (low energy/large scale) description of the
geometric properties of space-time and is experimentally very well tested.
However, the rapid development of observational cosmology during the last 20
years has led to data that cannot be explained by GR only. The most important of
these are the two phases of acceleration: inflation in the very early Universe
and the accelerated expansion of the Universe today. There are various attempts
to explain these two phases: cosmological constant, scalar field $\phi$, $f(R)$
and $f(T)$ theories and some other modifications of GR. 

In addition to these problems, no consistent (renormalizable) quantum field
theory of gravity has been constructed yet. Some candidates for a quantum
gravity are string theory, quantum loop gravity, dynamical
triangularization,\dots. Combining problems of divergences in quantum field
theory (QFT) and singularities in GR leads to discretized geometry
\cite{Doplicher}. Motivated by
quantum mechanics and Heisenberg uncertainty relations,  noncommutative (NC)
spaces can be defined \cite{NCbooks}. Then the nonzero commutation relations
between coordinates lead to discretization of space-time. Unfortunately, it is
not yet clear how to formulate a gravity theory on NC spaces (NC gravity) and
there are
various proposals in the literature. One can follow the twist approach in which
the commutative diffeomorphisms are replaced by the twisted diffeomorphisms
\cite{defgt}.
However, a full understanding of the twisted symmetries is still missing. Having
in mind that the NC gauge theories can be consistently defined, many authors
consider NC gravity as a gauge theory of the Lorentz/Poincar\' 
e group. These approaches are based on hermitian metrics \cite{Cha01} or
vielbeins \cite{ChPL,Cha04, PC09}. One can also construct emergent gravity from
noncommutative gauge theory and
matrix models, see \cite{Ste10}. Finally, there is the approach of NC
differential geometry and frame formalism \cite{MadoreMaja}.

Recently, a lot of attention has been given to the anti de Sitter (AdS)
gauge theory and to its application to GR
\cite{ChamMukh}, quantization of gravity \cite{Barrett}, AdS/CFT correspondence
and its
applications \cite{AdSCFT}. In our previous paper \cite{MiAdSGrav}, we begun the
study of noncommutative (NC) gravity based on the AdS gauge group. We started
with the
MacDowell-Mansouri action on the commutative space-time and generalized it to
the NC MacDowell-Mansouri action on the canonically deformed space. One of the
drawbacks of our approach was that we had to assume from the beginning that in
the commutative limit torsion vanishes. The other disadvantage was
that we introduced noncommutativity "in the middle": the symmetry breaking
from
$SO(2,3)$ to $SO(1,3)$ was performed in the
commutative model. The obtained $SO(1,3)$ invariant action was a basis for a
noncommutative gravity action. Using the enveloping algebra approach and the
Seiberg-Witten (SW) map \cite{jssw, Seiberg:1999vs} we constructed the NC
gravity action invariant under the NC $SO(1,3)_\star$ gauge
symmetry. The deformation of theory has not been introduced from the very
beginning, mostly for technical reasons: complicated calculations and gauge
non-invariant expressions. 

Nevertheless, it is of importance to have more general and more systematic
results. Therefore, in this
paper we analyze the full NC $SO(2,3)_\star$ gauge theory and perform symmetry
breaking after introducing the noncommutative deformation. The NC space-time we
work with is
the canonically deformed, with the Moyal-Weyl $\star$-product given by
\begin{equation}
\label{moyal}  f (x)\star  g (x) =
      e^{\frac{i}{2}\,\theta^{\a\b}\frac{\pa}{\pa x^\a}\frac{\pa}{ \pa
      y^\b}} f (x) g (y)|_{y\to x}\ .
\end{equation}
Here $\theta^{\mu\nu}$ is a constant antisymmetric matrix which is considered to
be a small deformation parameter. Indices $\mu$, $\nu$ take values $0,1,2,3$ and
the four dimensional Minkowski metric is $\eta_{\mu\nu} = diag(1,-1,-1,-1)$.  

In the next section we shortly describe the commutative $SO(2,3)$ gravity
theory as given in the literature and adapted to our notation.
In Section 3 the NC $SO(2,3)_\star$
gauge theory is constructed using the enveloping algebra approach and the SW
map. We then expand the NC action to the second
order in the deformation parameter $\theta^{\alpha\beta}$ and calculate
correction 
terms to the commutative action. We obtain that the first order
corrections vanish, thus we confirm the results already present in the
literature. The second order
correction is calculated using the method of composite fields developed in
\cite{PLM}. The correction terms we obtain are of zeroth to fourth power in the
curvature tensor and torsion. They are written in a manifestly covariant way.
However, the full result is very cumbersome and it is difficult to discuss its
physical implications. Fortunately, having three different scales in the model
enables us to discuss different limits of our model. To be able to compare
our results with $f(R)$ models present in the literature, in Section 4 we
analyze the limit of big cosmological constant and vanishing torsion and the limit of
small cosmological constant and vanishing torsion. In the limit of big cosmological
constant we obtain a $x$-dependent correction for the cosmological constant
and we analyze possible modifications of the zeroth order solution of vacuum
Einstein equations. Finally, we discuss the teleparallel limit, the limit in
which curvature vanishes and torsion is different from zero. Again we try to
compare our results with the existing results for $f(T)$ theories.

\section{Commutative gravity as AdS gauge theory}

\noindent In order to establish the notation, in this section we briefly review
the AdS gauge theory on four-dimensional Minkowski space-time. More details
about this construction can be found in \cite{MiAdSGrav}.

We assume that space-time has the structure of four-dimensional
Minkowski
space-time
$M_4$ and follow the usual steps for constructing a gauge theory on $M_4$ taking
the $SO(2,3)$ group as the gauge group. The gauge field is $SO(2,3)$-valued
\be
\omega_\m = \frac{1}{2}\omega_\m^{AB}M_{AB}\ ,\label{GaugePotAds}
\ee
with the generators of the $SO(2,3)$ group denoted by $M_{AB}$. The algebra is
given by
\be
[M_{AB},M_{CD}]=i(\eta_{AD}M_{BC}+\eta_{BC}M_{AD}-\eta_{AC}M_{BD}-\eta_{BD}M_{AC
})\ .
\label{AdSalgebra}
\ee
The $5 D$ metric is $\eta_{AB}={\rm diag}(+,-,-,-,+)$. Indices $A,B,\dots$ take
values
$0,1,2,3,5$, while indices $a,b,\dots$ take values $0,1,2,3$. A representation
of this algebra is given by
\bea
M_{ab} &=&\frac{i}{4}[\gamma_a,\gamma_b]=\frac12\sigma_{ab}\ ,\nonumber\\
M_{5a} &=&\frac{1}{2}\gamma_a\ , \label{Maba5}
\eea
where $\gamma_a$ are four dimensional Dirac gamma matrices. Then the gauge
potential $\omega_\m^{AB}$ decomposes into
$\omega_\m^{ab}$ and $\omega^{a5}_\m=\frac{1}{l} e_\mu^a$
\be
\omega_\m =
\frac{1}{2}\omega_\m^{AB}M_{AB}=\frac{1}{4}\omega_\m^{ab}\sigma_{ab}-
\frac{1}{2l} e_\m^{a}\gamma_a .\label{GaugePotAdsDecomp}
\ee
The parameter $l$ has dimension of length, while fields $e_\mu^a$ are
dimensionless. The meaning of the parameter $l$ will be clear at the end of
this section. Under the infinitesimal gauge transformations the gauge potential
transforms as
\be
\delta_\epsilon\omega_\mu=\pa_\mu\epsilon-i[\omega_\m,\epsilon],
\label{TrLawOmegaAB}
\ee
with the gauge parameter denoted by $\epsilon=\frac{1}{2}\epsilon^{AB}M_{AB}$.
The field
strength tensor is defined in the standard way as
\be
F_{\m\n}=\pa_\m\omega_\n-\pa_\n\omega_\m-i[\omega_\m,\omega_\n]
=\frac{1}{2}F^{AB}_{\mu\nu}M_{AB}
\ . \label{FAB}
\ee
Just like the gauge potential, the components of the field strength tensor
$F_{\m\n}^{\ AB}$ decompose into $F_{\m\n}^{\ ab}$ and $F_{\m\n}^{\ a5}$ . It is
easy to  show
that
\be
F_{\m\n}=\Big( R_{\m\n}^{\ ab}-\frac{1}{l^2}(e_\m^ae_\n^b-e_\m^be_\n^a)\Big)
\frac{\sigma_{ab}}{4} - F_{\m\n}^{\ a5}\frac{\gamma_a}{2}\ , \label{FabFa5}
\ee
where
\bea
R_{\m\n}^{\ ab} &=&
\pa_\m\omega_\n^{ab}-\pa_\n\omega_\m^{ab}+\omega_\m^{ac}\omega_\n^{cb}
-\omega_\m^{bc}\omega_\n^{ca} \label{Rab}\\
lF_{\m\n}^{\ a5} &=& D_\m e^a_\n-D_\n e^a_\m = T_{\m\n}^a .\label{Ta}
\eea
Under the infinitesimal gauge transformation the field strength transforms
covariantly
\be
\delta_\epsilon F_{\m\n}=i[\epsilon, F_{\m\n}]. \label{TrLawFAB}
\ee

Equations (\ref{GaugePotAdsDecomp}), (\ref{FabFa5}), (\ref{Rab}) and
(\ref{Ta}) suggest that one can identify $\omega^{ab}_\mu$
with the spin connection of the Poincar\' e gauge theory, $e^{a}_\mu$ with
the
vierbeins, $R^{\ \ ab}_{\mu\nu}$ with the curvature tensor and $lF^{a5}_{\mu\nu}$
with
the torsion.

Indeed, it was shown in the seventies that one can do such
identification and relate AdS
gauge theory with GR. Different ways were discussed in the literature, see
\cite{ stelle-west, McD-Mansouri, Towsend}. One way is to start from the
action which contains a scalar field
\be
S = \frac{il}{64\pi G_N}\tr \int{\rm d}^4x \epsilon^{\mu\nu\rho\sigma}
F_{\mu\nu} F_{\rho\sigma}\phi \ ,\label{KomDejstvo}
\ee
where $G_N$ is the Newton gravitational constant. The auxiliary field
$\phi=\phi^A \Gamma_A$, $\Gamma_A =(i\gamma_a\gamma_5, \gamma_5)$ transforms
in the adjoint representation of $SO(2,3)$
\be
\delta \phi = i[\epsilon,\phi]\ . \label{DeltaPhi}
\ee
The action (\ref{KomDejstvo}) is invariant under the $SO(2,3)$ gauge symmetry.
However, if we break the symmetry and restrict the field $\phi$ to be
$\phi^a=0,\
\phi^5=l$ then the symmetry of the action is reduced to the $SO(1,3)$ gauge
symmetry. The constraint on the field $\phi$ can be introduced in various
ways via a Lagrange
multiplier or dynamically \cite{stelle-west}. The action obtained after
symmetry breaking is then given by
\bea
S &=& \frac{il^2}{64\pi G_N}\epsilon^{\m\n\r\s}\int {\rm d}^4 x\tr(F_{\m\n}
F_{\r\s}\gamma_5)\nn \\
&=& -\frac{1}{16\pi G_N}\int {\rm d}^4 x\Big[\frac{l^2}{16}\epsilon^{\m\n\r\s}
\epsilon_{abcd}R_{\m\n}^{\ ab}R_{\r\s}^{\ cd} + eR + 2e\Lambda \Big]\ ,
\label{com.dejstvo.triclana}\eea
where $\Lambda =-3/l^2$ and $e=\det (e_\m^a)$. In the first line we inserted
expansions
(\ref{FabFa5}) and (\ref{Rab}) and after some standard manipulation with indices
and traces we obtained the
second line. The action (\ref{com.dejstvo.triclana}) appeared for the first time
in the paper by MacDowell and Mansouri
\cite{McD-Mansouri}.

This action is written in the first order formalism: the spin connection
$\omega^{ab}_\mu$ and the vierbeins $e_\mu^a$ are independent fields. The
corresponding equations of motion give vanishing torsion and enable to
express the spin connection as a function of vierbeins. Inserting the
solution for the spin connection in the action (\ref{com.dejstvo.triclana})
gives the action in the second order formalism: the only dynamical (propagating)
field is the metric $g_{\mu\nu} = \eta_{ab}e_\mu^a e_\nu^b$. In that case the
first term in (\ref{com.dejstvo.triclana})
is the Gauss-Bonnet term; it is a topological term and does not contribute to
the
equations of motion. The second term is the Einstein-Hilbert action, while the
last term is
the cosmological constant term. Therefore,
after symmetry breaking the action (\ref{KomDejstvo}) describes GR with the
negative cosmological constant and the topological Gauss-Bonnet term. We see
that the parameter $l$ is related with the cosmological constant and the radius
of AdS space-time. AdS space is a solution of Einstein vacuum equations
obtained from this action.

\section{The NC $SO(2,3)_\star$ gauge theory}

\noindent In this section we generalize the model (\ref{KomDejstvo}) to the
noncommutative case. We work with the simplest form of noncommutativity,
canonical or
$\theta$-constant noncommutativity. Following the approach
of deformation quantization we represent noncommutative functions as functions
of
commuting coordinates and algebra multiplication with the Moyal-Weyl
$\star$-product (\ref{moyal}). The noncommutativity (deformation) is encoded in
the $\star$-product, while all
variables (fields) are functions of commuting coordinates. Integration is well
defined since the usual integral is cyclic:
\be
\int {\rm d}^4 x (f\star g\star h ) = \int {\rm d}^4 x ( h\star f\star g )\
 + {\mbox{boundary terms}}.\label{cyclicity}
\ee
In general, depending on the behavior of fields at the boundary,
these boundary terms can be different from zero. The boundary terms do not
influence equations of motion, but might be needed to have a well defined
variational principle, that is a well defined functional derivative of the
action\footnote{The functional variation of a general action for a filed
$\phi$ can be given by
$$
\delta S = \int{\rm d}^4 x \Big( E\delta \phi + \partial_\mu(B_1\delta \phi)
+\partial_\mu (B_2\delta\partial_\mu \phi) \Big) \, .
$$
The first term gives the equations of motion $E=0$. The term with $B_1$ is a
surface term and
it vanishes since the variation of the filed $\phi$ is zero at the boundary.
The term with $B_2$ is again a surface term. However, it does not vanishes,
since the
variation of $\partial_\mu\phi$ does not have to be zero at the boundary. To
cancel this term, one adds a boundary term to the starting action. Then the new
action has a well defined functional derivative. This situation is typical in
gravity theories on commutative space-time.}. For example, in the case of
Einstein-Hilbert action the necessary boundary term is the Gibbons-Hawking-York
term \cite{GHY, Wald}. In this paper we
calculate the NC gravity action up to second order in the
deformation parameter and the equations of motion which follow from it.
Therefore, in the following we will omit boundary terms. Their analysis in the
noncommutative theories is nontrivial. It is a subject for itself and we
postpone it for future work.

In particular, from (\ref{cyclicity}) we have $\int {\rm d}^4 x (f\star g) = \int {\rm d}^4 x (g\star f) =\int
{\rm d}^4 x fg$. Note that the volume element ${\rm d}^4 x$ is not $\star$-multiplied
with the functions under the integral.

\subsection{The Seiberg-Witten map}

\noindent In order to construct the NC $SO(2,3)_\star$ gauge theory we use the
enveloping algebra approach
and the Seiberg-Witten map developed in \cite{jssw, Seiberg:1999vs}. Under the
infinitesimal
NC $SO(2,3)_\star$ gauge transformations the NC gauge field $\hat{\omega}_\mu$
transforms as
\begin{equation}
\delta_\epsilon^\star{\hat\omega}_\m = \pa_\m{\hat\Lambda}_\epsilon
+ i[ {\hat\Lambda}_\epsilon\ds {\hat\omega}_\m] ,\label{SO23Omega}
\end{equation}
with the NC gauge parameter $\hat{\Lambda}_\epsilon$. In the commutative limit
$\hat{\Lambda}_\epsilon$ reduces to the commutative gauge parameter $\epsilon
=\frac{1}{2}\epsilon^{AB}M_{AB}$. We demand consistency of the NC gauge transformations
\be
[\delta_{\epsilon_1}^\star\ds\delta_{\epsilon_2}^\star]=\delta_{-i[\epsilon_1,
\epsilon_2]}^
\star\ .\label{SO23GaugeAlgebra}
\ee
This will be the case provided that the gauge parameter $\hat{\Lambda}_\epsilon$
is in the enveloping algebra of
the $so(2,3)$ algebra\footnote{Note that in (\ref{SO23Omega})
$\star$-commutators appear. These commutators do not close in the Lie algebra.
Namely, if $A=A^aT^a$ and $B=B^aT^a$ then
$$
[A\ds B] = \frac{1}{2}(A^a\star B^b + B^b\star A^a)[T^a,T^b] +
\frac{1}{2}(A^a\star B^b - B^b\star A^a)\{T^a,T^b\} .
$$
Only in the case of $U(N)$ in the fundamental representation the anticommutator
of generators is still in the corresponding Lie algebra.}. However, an enveloping
algebra is infinitely dimensional and the resulting theory seems to have
infinitely many degrees of freedom. This problem is solved by the Seiberg-Witten
map. The idea of the Seiberg-Witten map is that the NC gauge transformations are
induced by the corresponding commutative gauge transformations
\be
{\hat\omega}_\m(\omega)+\delta_\epsilon^\star{\hat\omega}_\m(\omega)={\hat\omega
}_\m(\omega+
\delta_\epsilon\omega) \ ,\label{DeltaOmegaSO23}
\ee
with the commutative gauge field $\omega_\mu$ and the commutative gauge
parameter
$\epsilon$. As a result of this, all noncommutative
variables (gauge parameter, fields) can be expressed in terms of the
corresponding
commutative variables and their derivatives as power series in the
noncommutativity
parameter $\theta^{\alpha\beta}$. 

In the case of NC gauge parameter the expansion is
\be
\hat{\Lambda}_\epsilon = \Lambda^{(0)} + \Lambda^{(1)} + \Lambda^{(2)}\dots
.\nonumber
\ee
Inserting this expansion into (\ref{SO23GaugeAlgebra}) and expanding all
$\star$-products gives a variational equation for the NC gauge parameter. This
equation can be solved to all orders of the deformation parameter. The zeroth
order solution is the commutative gauge parameter $\epsilon$, as mentioned
earlier. The recursive relation between the $n$th and the $(n+1)$st order
solution is given by \cite{kayahan}, \cite{PC11}
\be 
{\hat\L}^{(n+1)} = -\frac{1}{4(n+1)}\theta^{\kappa\lambda}\Big( \{\hat
{\omega}_\kappa \ds
\pa_\lambda {\hat \epsilon} \}\Big)^{(n)} \ , \label{RecRelLambda}
\ee
where $(A\star B)^{(n)} = A^{(n)}B^{(0)} + A^{(n-1)}B^{(1)} + \dots
+ A^{(0)}\star ^{(1)} B^{(n-1)} + A^{(1)}\star ^{(1)} B^{(n-2)} +\dots$ includes
all
possible terms of order $n$. Expanding this recursive relation we obtain
\bea
\hat{\Lambda} &=& \epsilon -\frac{1}{4}\theta^{\alpha\beta}\{\omega_\alpha,
\partial_\beta\epsilon
\} +{\cal O}(\theta^2) \label{SO23Lambda1}\\
&=& \frac{1}{2}\Lambda^{AB}M_{AB} + \Lambda^A \Gamma_A + \Lambda I \nonumber\\
&=& \frac{1}{4}\Lambda^{ab}\sigma_{ab} + \Lambda^a \gamma_a + \tilde{\Lambda}^a
\gamma_a\gamma_5 + \tilde{\Lambda}\gamma_5 + \Lambda I \ .\label{UEALambda}
\eea
For example, $\Lambda^{AB (0)} = \epsilon^{AB}$, while $\Lambda^{A (0)} =0$.
From the first line it is obvious that $\hat{\Lambda}$ is enveloping algebra
valued\footnote{The advantage of working with the $\gamma$-matrix representation
is that the enveloping algebra is finite.}.

Solving the equation
\be
{\hat\omega}_\m(\omega)+\delta_\epsilon^\star{\hat\omega}_\m(\omega)={\hat\omega
}_\m(\omega+
\delta_\epsilon\omega) \label{DeltaOmegaSO13}
\ee
order by order in the NC parameter the noncommutative gauge
field $\hat{\omega}_\m$ is expressed in terms of the commutative gauge field
$\omega_\mu$.
The recursive relation in this case is given by
\be
{\hat\omega}_\m^{(n+1)}= -\frac{1}{4(n+1)}\theta^{\kappa\lambda}
\Big( \{{\hat \omega}_\kappa \ds \pa_\lambda{\hat \omega}_\m + {\hat F}_{\l\m}\}
\Big)^{(n)} .\label{RecRelOmega}
\ee
The gauge field ${\hat\omega}_\m$ is of the form
\bea
{\hat\omega}_\m &=& \omega_\mu
-\frac{1}{4}\theta^{\kappa\lambda}\{\omega_\kappa, (\partial_\lambda\omega_\mu +
F_{\lambda\mu}\}
+ {\cal O}(\theta^2) \label{SO23Omega1}\\
&=& \frac{1}{4}\omega^{ab}_\m\sigma_{ab}  + \omega_\mu^a \gamma_a
+\tilde{\omega}_\mu^a\gamma_a\gamma_5 
+ {\tilde\omega}^5_\m\gamma_5 + \omega_\m I \ . \label{UEAOmega}
\eea
The NC field strength tensor is defined as
\be
{\hat F}_{\m\n}=\pa_\m{\hat \omega}_\n-\pa_\n{\hat \omega}_\m
-i[{\hat\omega}_\m\ds {\hat\omega}_\n] \label{nckrivina}
\ee
and 
\begin{equation}
\delta^\star_\epsilon {\hat F}_{\m\n}= i[\hat{\Lambda}_\epsilon \ds {\hat
F}_{\m\n}]\ .\label{DeltaFStar}
\end{equation}
The SW map solution for ${\hat F}_{\mu\nu}$ follows from
the definition (\ref{nckrivina}), using the result (\ref{RecRelOmega}). The
recursive formula is
\bea
{\hat F}_{\m\n}^{(n+1)} &=& -\frac{1}{4(n+1)}\theta^{\kappa\lambda}\Big( \{
{\hat \omega}_\kappa \ds
\partial_\lambda {\hat F}_{\mu\nu} + D_\lambda {\hat F}_{\mu\nu} \} \Big)^{(n)}
\nn\\
&& +\frac{1}{2(n+1)}\theta^{\kappa\lambda}\Big( \{ {\hat F}_{\mu\kappa}, \ds
{\hat F}_{\nu\lambda} \}
\Big)^{(n)} \ .\label{RecRelR}
\eea
Note that we do not put a "hat" on the covariant derivative $D_\mu$, the meaning
of $D_\mu$ is defined by the expression it acts on: $D_\lambda \hat{F}_{\m\n} =
\partial_\lambda \hat{F}_{\mu\nu}
-i[\hat{\omega}_\lambda \ds \hat{F}_{\mu\nu}]$ and $D_\lambda F_{\m\n} =
\partial_\lambda F_{\mu\nu}
-i[\omega_\lambda, F_{\mu\nu}]$. One can check that
\bea
{\hat F}_{\mu\nu} &=& F_{\mu\nu} -\frac{1}{4}\theta^{\kappa\lambda}
\{\omega_\kappa,\pa_\lambda F_{\m\n} + D_\lambda F_{\m\n}\} +
\frac{1}{2}\theta^{\kappa\lambda} \{F_{\mu\kappa}, F_{\nu\lambda} \}
+ {\cal O}(\theta^2) \label{SO23F1}\\
&=& \frac{1}{4} F^{\ ab}_{\m\n}\sigma_{ab} + F^a\gamma_a +
\tilde{F}^a\gamma_a\gamma_5
+ {\tilde  F}^5_{\m\n}\gamma_5  + F_{\m\n} I+{\cal O}(\theta^2) . \label{UEAF}
\eea

Finally, the field $\hat\phi$ transforms in the adjoint representation 
\be
\delta_\epsilon^\star{\hat \phi} = i[{\hat\Lambda}_\epsilon\ds{\hat \phi}]\  .
\label{DeltaPhiStar}
\ee
Using the previous results we find the recursive relation
\be
\hat{\phi}^{(n+1)} = -\frac{1}{4(n+1)}\theta^{\kappa\lambda} \Big( \{{\hat
\omega}_\kappa \ds \pa_\l {\hat{\phi}} + D_\l {\hat{\phi}} \} \Big)^{(n)} \ ,
\label{RecRelPhi}
\ee
with $D_\l {\hat \phi} = \partial_\l {\hat \phi} -i [{\hat \omega}_\l \ds {\hat \phi} ]$ and $D_\l \phi = \partial_\l \phi -i [\omega_\l ,\phi ]$.
The solution for $\hat{\phi}$ has the following structure
\bea
{\hat \phi} &=& \phi -\frac{1}{4}\theta^{\kappa\lambda}
\{\omega_\kappa,\pa_\lambda\phi + D_\lambda \phi\} + {\cal O}(\theta^2)
\label{SO23Phi1}\\
&=& \phi^a\gamma_a\gamma_5 + \phi\gamma_5 + \frac{1}{4}\phi^{ab}\sigma_{ab} +
\tilde{\phi}^a\gamma_a+{\cal O}(\theta^2)\ . \label{UEAPhi}
\eea
Note that a term proportional to the unit matrix is absent from the first
order solution. This is a consequence of the algebra of $M_{AB}$ and $\Gamma_A$
matrices, see the list of identities in Appendix. The term proportional to the
unit matrix will appear in the second and higher orders.

\subsection{The NC AdS action}

\noindent The NC action is now given by
\be
S_{NC} = \frac{il}{64\pi G_N}\tr \int{\rm d}^4x \epsilon^{\mu\nu\rho\sigma}
\hat{F}_{\mu\nu}\star \hat{F}_{\rho\sigma}\star \hat{\phi}\, .\label{NCdejstvo}
\ee
The $\star$-product is the Moyal-Weyl $\star$-product (\ref{moyal}),
fields with a "hat" are NC fields and we will use the SW map solutions
(\ref{SO23F1}), (\ref{SO23Phi1}). Using the transformation laws
(\ref{DeltaFStar}), (\ref{DeltaPhiStar}) and the cyclicity of the integral
(\ref{cyclicity}) one can show that this action is invariant under the NC
$SO(2,3)_\star$ gauge transformations\footnote{The NC gauge variation of the
action (\ref{NCdejstvo}) is given by
$$
\delta_\epsilon^\star S_{NC} \sim \int {\rm d}^4 x
\epsilon^{\mu\nu\rho\sigma}{\rm Tr} [\Lambda_\epsilon \ds \hat{F}_{\mu\nu}\star
\hat{F}_{\rho\sigma}\star \hat{\phi}] = \oint {\rm d}\Sigma_\mu K^\mu
$$
where $K^\mu$ is a function of $\varepsilon$, $\omega_\mu$, $F_{\mu\nu}$, their
derivatives
and $\theta^{\a\b}$. This is not unusual in the field theory. For example, the
variation of the commutative Einstein-Hilbert action under infinitesimal
diffeomorphisms generated by the vector field $\xi=\xi^\mu\partial_\mu$ is
different form zero and is given by a surface term. This surface term vanishes
if one demands that $\xi^\mu=0$ at the boundary. This is a standard textbook
procedure, see \cite{Peskin} and references therein. Therefore, if we demand
that the gauge parameter $\varepsilon$ has appropriate behavior at the
boundary, the surface term vanishes and the variation of the action
(\ref{NCdejstvo}) is zero. A
detailed analysis of boundary terms is required if one discussed conserved
quantities, especially in gravity, see Chapter 7 in \cite{MilutinKnjiga}.}. In
the limit $\theta^{\alpha\beta}\to 0$
the
action (\ref{NCdejstvo}) reduces to the commutative action (\ref{KomDejstvo}).

Let us now calculate the first order correction to (\ref{KomDejstvo}). To
this end, we calculate
\bea
({\hat F}_{\m\n}\star{\hat F}_{\r\s})^{(1)}&=& F_{\m\n}^{(1)} F_{\r\s}
+F_{\m\n}F_{\r\s}^{(1)} +\frac{i}{2}\theta^{\a\b}\pa_\a F_{\m\n}\pa_b
F_{\r\s}\nn\\
&=&-\frac14\theta^{\a\b}\{\omega_\a,\pa_\b(F_{\m\n}F_{\r\s}) + D_\b(F_{
\m\n } F_{\r\s})\}\nn\\
&& +\frac{i}{2}\theta^{\a\b}(D_\a
F_{\m\n})(D_{\b}F_{\r\s})+\frac{1}{2}\theta^{\a\b}(\{
F_{\a\m},F_{\b\n}\}F_{\r\s}\nn\\
&& +F_{\m\n}\{F_{\a\r,F_{\b\s}}\}) \label{FF1}
\eea
and
\bea 
({\hat F}_{\m\n}\star{\hat F}_{\r\s} \star \hat{\phi})^{(1)}&=&({\hat F}_{\m\n}\star{\hat F}_{\r\s})^{(1)}\phi + (F_{\m\n}
F_{\r\s})\phi^{(1)} + \frac{i}{2}\theta^{\a\b}\pa_\a(F_{\m\n}
F_{\r\s})\pa_\b\phi\nn\\
&=&-\frac14\theta^{\a\b}\{\omega_\a,(\pa_\b+D_\b)(F_{\m\n}F_{\r\s}\phi)\}
\label{FFPhi1} \\
&& +\frac{i}{2}\theta^{\a\b}D_\a (F_{\m\n}F_{\r\s})D_\b\phi +
\frac{i}{2}\theta^{\a\b}D_\a F_{\m\n}D_\b
F_{\r\s}\phi\nn\\
&& +\frac12\theta^{\a\b}\{F_{\a\m},F_{\b\n}\}
F_{\r\s}\phi+\frac12\theta^{\a\b}F_{\m\n}\{F_{\a\r},F_{\b\s}\}\phi \ . \nn
\eea
Then, the first order of the NC action (\ref{NCdejstvo}) is given by
\bea 
S_{NC}^{(1)}& =& \frac{il}{64\pi G_N}\tr \int{\rm d}^4x
\epsilon^{\mu\nu\rho\sigma}
(\hat{F}_{\mu\nu}\star \hat{F}_{\rho\sigma}\star \hat{\phi})^{(1)}\nn\\
&=&\frac{il}{64\pi G_N}\theta^{\a\b}\tr \int{\rm d}^4x
\epsilon^{\mu\nu\rho\sigma}\Big( -\frac14F_{\m\n}F_{\r\s}\{F_{\a\b},\phi\}\nn\\
&& +\frac{i}{2}D_\a F_{\m\n}D_\b
F_{\r\s}\phi \label{NCDejstvo1} \\
&& +\frac12\{F_{\a\m},F_{\b\n}\}
F_{\r\s}\phi + \frac12F_{\m\n}\{F_{\a\r},F_{\b\s}\}\phi
\Big) \ . \nn
\eea
We performed one partial integration to obtain (\ref{NCDejstvo1}) from
(\ref{FFPhi1}). After explicitly calculating the traces of the products of
gamma
matrices, we obtain $S_{NC}^{(1)}=0$. In this way, once again we confirm the
statement already present in the literature: if the reality of action is
imposed, then there is no first order correction
for the NC gravity action. This results seems to be model independent
\cite{ChPL, PC09, Mukherjee2}.

Similarly to the SW map solutions, there is a recursive relation between the
$(n+1)$st and the $n$th order of the expansion of (\ref{NCdejstvo}). In
particular, the second order correction is given by
\bea 
S_{NC}^{(2)} &=& \frac{il}{128\pi G_N}\theta^{\a\b}\tr \int{\rm d}^4x
\epsilon^{\mu\nu\rho\sigma}
\Big(
-\frac14\hat{F}_{\m\n}\star\hat{F}_{\r\s}\star\{\hat{F}_{\a\b}\ds\hat{\phi}\}
\nn\\
&& +\frac{i}{2}D_\a {\hat F}_{\m\n}\star D_\b
\hat{F}_{\r\s}\star \phi \label{NCDejstvo2}\\ 
&& +\frac12\{\hat{F}_{\a\m}\ds\hat{F}_{\b\n } \}\star
\hat{F}_{\r\s}\star\hat\phi+\frac12\hat{F}_{\m\n}\star\{\hat{F}_{
\a\r } \ds \hat {F}_ {\b\s} \}\star \hat\phi
\Big)^{(1)} \ .\nn
\eea
In order to calculate this expression, we have to expand the $\star$-products in
(\ref{NCDejstvo2}) and use the SW map solutions (\ref{SO23F1}) and
(\ref{SO23Phi1}). However, inserting these solutions straightforwardly into
(\ref{NCDejstvo2}) gives lots of noncovariant terms: terms with partial
derivatives and the "naked" gauge field $\omega_\mu$. To avoid this, we use the
method of composite fields \cite{PLM} which enables to write the result in a
manifestly gauge covariant way. The first, third and fourth term can be
calculated similarly to what has been done in (\ref{FF1}) and (\ref{FFPhi1}) and
we will not go into
details here. The useful formula is
\begin{eqnarray}
&&\hat{(F_{\m\n}}\star \hat{F}_{\r\s}\star \hat{F}_{\a\b}\star\hat{\phi})^{(1)} =
-\frac14\theta^{\k\l}\{\omega_\k,(\pa_\l+D_\l)(F_{\m\n}F_{\r\s}F_{\a\b}\phi)\}
\label{FFFPhi1} \\
&& +\frac{i}{2}\theta^{\k\l}D_\k F_{\m\n}D_\l(F_{\r\s}F_{a\b}\phi) +
\frac{i}{2}\theta^{\k\l}F_{\m\n}\Big( D_\k (F_{\r\s} F_{\a\b})D_\l \phi +  D_\k
F_{\r\s} D_\l F_{\a\b} \phi\Big) \nn\\
&& +\frac12\theta^{\k\l} \Big( \{F_{\k\m},F_{\l\n}\} F_{\r\s}F_{\a\b} +
F_{\m\n}\{F_{\k\r},F_{\l\s}\} F_{\a\b} + F_{\m\n}F_{\r\s} \{ F_{\k\a},
F_{\l\b}\} \Big)\phi \ . \nn
\end{eqnarray}
The second term containing the covariant
derivative $D_\mu$ we calculate in details. Starting from
\bea
&&(D_\a\hat{F}_{\mu\nu})^{(1)} =
-\frac14\theta^{\kappa\l}\{\omega_\kappa,\pa_\l(D_\a F_{\mu\nu})
+D_\l(D_\a F_{\mu\nu} )\} + \frac{1}{2}\theta^{\k\l}\{F_{\k\a}, D_\l F_{\m\n}\}
\nn\\
&& + \frac{1}{2}\theta^{\k\l}\{ D_\a F_{\k\m}, F_{\l\n} \} +
\frac{1}{2}\theta^{\k\l}\{ F_{\k\m}, D_\a F_{\l\n} \} \ ,\label{DF}
\eea
we obtain
\bea
&&(D_\a\hat{F_{\mu\nu}}\star D_\b\hat{F_{\rho\sigma}})^{(1)} =
-\frac14\theta^{\kappa\l}\{\omega_\kappa,
\pa_\l(D_\a F_{\mu\nu} D_\b F_{\rho\sigma}) + D_\l(D_\a F_{\mu\nu} D_\b
F_{\rho\sigma})\}\nn\\
&& + \frac{i}{2}\theta^{\k\l}(D_\k D_\a F_{\m\n})(D_\l D_\b F_{\r\s})
\label{DFDF}\\
&& + \frac{1}{2}\theta^{\k\l} \Big( \{F_{\k\a}, D_\l F_{\m\n}\} + \{ D_\a
F_{\k\m}, F_{\l\n} \} + \{ F_{\k\m}, D_\a F_{\l\n} \}\Big)(D_\b F_{\r\s})\nn\\
&& + \frac{1}{2}\theta^{\k\l} (D_\a F_{\m\n})\Big( \{F_{\k\b}, D_\l F_{\r\s}\} +
\{ D_\b F_{\k\r}, F_{\l\s} \} + \{ F_{\k\r}, D_\b F_{\l\s} \}\Big) \ .\nn
\eea
We used that $D_\alpha F_{\mu\nu}$ is a field in the adjoint
representation. This gives the first term on the RHS of (\ref{DF}). Then the
remaining terms of SW map solutions for $\hat{F}_{\mu\nu}$ and
$\hat{\omega}_\mu$ are arranged in a gauge covariant way. To calculate
(\ref{DFDF})
we again used that $D_\a\hat{F_{\mu\nu}}\star D_\b\hat{F_{\rho\sigma}}$ is a
field in the adjoint representation. That gives the first term
on the RHS of (\ref{DFDF})). The remaining terms have to be covariant. The second
term on the RHS is the covariant $\star$-product and the additional terms come
from the covariant terms in (\ref{DF}) and (\ref{SO23F1}). Applying similar
steps finally leads to 
\bea
&&(D_\a\hat{F_{\mu\nu}}\star D_\b\hat{F_{\rho\sigma}}\star
\hat{\phi})^{(1)}=-\frac14\theta^{\kappa\l}\{\omega_\kappa,
\pa_\l(D_\a F_{\mu\nu} D_\b F_{\rho\sigma}\phi) + D_\l(D_\a F_{\mu\nu} D_\b
F_{\rho\sigma}\phi)  \nn\\
&& + \frac{i}{2}\theta^{\k\l}\Big( D_\k (D_\a F_{\mu\nu}D_\b
F_{\rho\sigma})(D_\l\phi) +(D_\k D_\a F_{\m\n})(D_\l D_\b F_{\r\s})\phi\Big)
\nn\\
&& + \frac{1}{2}\theta^{\k\l}\Big( \big(\{F_{\k\a}, D_\l F_{\m\n}\} + \{ D_\a
F_{\k\m}, F_{\l\n} \} + \{ F_{\k\m}, D_\a F_{\l\n} \}\big)(D_\b F_{\r\s})
\nn\\
&& + (D_\a F_{\m\n})\big( \{F_{\k\b}, D_\l F_{\r\s}\} + \{ D_\b F_{\k\r},
F_{\l\s} \} + \{ F_{\k\r}, D_\b F_{\l\s} \}\big) \Big)\phi \ .
\label{DFDFPhi}
\eea
Collecting the results for all three terms in
(\ref{NCDejstvo2}) we obtain
\bea 
S_{NC}^{(2)} &=& \frac{il}{64\pi
G_N}\frac18\theta^{\a\b}\theta^{\kappa\l}\tr \int{\rm d}^4x
\epsilon^{\mu\nu\rho\sigma}\Big\{ \frac18\{
F_{\a\b},\{F_{\m\n},F_{\r\s}\}\}\{\phi,F_{\kappa\l}\} \nn\\
&& -\frac12\{F_{\a\b},\{F_{\r\s},\{F_{\kappa\m},F_{\l\n}\}\}\}\phi
-\frac14\{\{F_{\m\n},F_{\r\s}\},\{F_{\kappa\a},F_{\l\b}\}\}\phi \nn\\
&& -\frac{i}{4}\{F_{\a\b},[D_\kappa
F_{\m\n},D_{\l}F_{\r\s}]\}\phi - \frac{i}{2}[\{D_\kappa
F_{\m\n},F_{\r\s}\},D_{\l}F_{\a\b}]\phi \nn\\
&& -\frac12\{
F_{\r\s},\{F_{\a\m},F_{\b\n}\}\}\{\phi,F_{\kappa\l}\}
+\{ \{F_{\a\m},F_{\b\n} \},\{F_{\kappa\r},F_{\l\s}\}\}\phi\nn\\
&& +2\{F_{\r\s},\{F_{\b\n},\{F_{\kappa\a},F_{\l\m}\}\}\}\phi
+ i\{F_{\r\s},[D_\kappa F_{\a\m},D_\l F_{\b\n}]\}\phi \nn\\
&& +2i[\{F_{\b\n},D_{\kappa}F_{\a\m}\},D_\l F_{\r\s}]\phi \nn\\
&&-\frac{i}{4}\{\phi,F_{\kappa\l}\}[D_\a F_{\m\n},D_\b F_{\r\s}]
-\frac{1}{2}\{D_\kappa D_\a F_{\m\n},D_\l D_\b F_{\r\s}\}\phi\nn\\
&& + i[\{ F_{\k\a},D_\l F_{\m\n}\},D_\b F_{\r\s}]\phi
+ i[\{ F_{\l\n},D_\a F_{\kappa\m}\},D_\b F_{\r\s}]\phi \nn\\
&& + i[\{ F_{\k\m},D_\a F_{\l\n}\},D_\b F_{\r\s}]\phi\Big\} \ .
\label{NCDejstvo2Exp}
\eea
This expanded action is obviously invariant under the commutative $SO(2,3)$
gauge transformations, as guaranteed by the SW map. In order to break this
symmetry down to $SO(1,3)$ we have to constrain the commutative field $\phi$ to
be of the form $\phi=(0,0,0,0,l)$. In this way, in the limit $\theta^{\a\b}\to
0$ we
obtain the commutative action (\ref{com.dejstvo.triclana}). The second order
correction after the symmetry breaking and after calculating traces is given
by
\bea 
S_{NC}^{(2)} &=& -\frac{l^2}{64\pi
G_N}\theta^{\a\b}\theta^{\kappa\l}
\epsilon^{\mu\nu\rho\sigma}\epsilon_{abcd} \int{\rm
d}^4x\Big\{ \frac{1}{256}\Big(F_{\m\n}^{\ \ cd}
F_{\r\s}^{\ \ ab}F_{\a\b}^{\ \ mn}F_{\kappa\l mn} \nn\\
&&-8F_{\m\n}^{\ \
ab}F_{\r\s}^{\ \ c5}F_{\kappa\l}^{\ \ de}F_{\a\b e}^{\ \ \ 5}
+F_{\a\b}^{\ \ ab}F_{\kappa\l}^{\ \ cd}(F_{\m\n}^{\ \ mn}F_{\r\s
mn}+2F_{\m\n}^{\ \ m5}F_{\r\s m}^{\ \ \ 5})
\Big)\nn\\
&&-\frac{1}{32}\Big(F_{\kappa\l}^{\ \ ab}F_{\m\n}^{\ \ cd}F_{\a\r}^{\ \
mn}F_{\b\s mn}+2F_{\a\b}^{\ \ ab}F_{\r\s}^{\ \ cd}F_{\kappa\m}^{\ \
m5}F_{\l\n m }^{\ \ \ 5} \nn\\
&&+F_{\kappa\m}^{\ \ ab}F_{\l\n}^{\ \ cd}F_{\a\b}^{\ \
mn}F_{\r\s mn}\Big)\nn\\
&&-\frac{1}{128}\Big(F_{\kappa\a}^{\ \ ab}F_{\l\b}^{\ \ cd}(F_{\m\n}^{\ \
mn}F_{\r\s mn}\nn\\
&&+2F_{\m\n}^{\ \ m5}F_{\r\s m}^{\ \ \ 5})+ F_{\m\n}^{\ \
ab}F_{\r\s}^{\ \ cd}(F_{\kappa\a}^{\ \ mn}F_{\l\b mn}+2F_{\kappa\a}^{\ \
m5}F_{\l\b m}^{\ \ \ 5}) \Big)\nn\\
&&+\frac{1}{16}F_{\a\b}^{\ \ ab}\Big((D_\kappa F_{\m\n})^{cm})(D_\l
F_{\r\s})^{\ d}_m
+(D_\kappa F_{\m\n})^{c5}(D_\l
F_{\r\s})^{\ d}_5\Big)\nn\\
&&-\frac{1}{16}\Big( (D_\kappa F_{\m\n})^{ab}(D_\l F_{\a\b})^{d5}F_{\r\s}^{c5}
+(D_\kappa F_{\m\n})^{a5}(D_\l F_{\a\b})^{b5}F_{\r\s}^{\ \ cd} \Big)\nn\\
&&+\frac{1}{16}F_{\a\m}^{ab}F_{\b\n}^{cd}\Big(F_{\kappa\r}^{\ \
mn}F_{\l\s mn}+2F_{\kappa\r}^{\ \ m5}F_{\l\s m5}\Big)\nn\\
&&+\frac{1}{16}\Big(F_{\r\s}^{\ \ ab} F_{\b\n}^{\ \ cd}(F_{\kappa\a}^{\ \
mn}F_{\l\m mn}+2F_{\kappa\a}^{\ \ m5}F_{\l\m m5})+F_{\kappa\a}^{\ \
ab}F_{\l\m}^{\ \ cd}F_{\r\s}^{\ \ mn}F_{\b\n mn}\nn\\
&&-4(F_{\kappa\a}^{\ \ ab}F_{\l\m}^{\ \ c5}+F_{\kappa\a}^{\ \ a5}F_{\l\m}^{\ \
bc})F_{\r\s}^{\ \ de}F_{\b\n  e5} \Big)\nn\\
&&-\frac18F_{\r\s}^{\ \ ab}\Big((D_\kappa F_{\a\m})^{cm}(D_\l
F_{\b\n})_m^{ \ d}+(D_\kappa F_{\a\m})^{c5}(D_\l
F_{\b\n })_5^{ \ d}\Big)\nn\\
&&+\frac12\Big(F_{\kappa\m}^{\ \ ab}(D_\a F_{\l\n})^{c5}(D_\b
F_{\r\s})^{d5}+F_{\kappa\m}^{\ \ a5}(D_\a F_{\l\n})^{bc}\Big)(D_\b
F_{\r\s})^{d5}\nn\\
&&+\frac18\Big(F_{\kappa\a}^{\ \ ab}(D_\l
F_{\m\n})^{c5}+F_{\kappa\a}^{\ \ a5}(D_\l F_{\m\n})^{bc}\Big)(D_\b
F_{\r\s})^{d5}\nn\\
&&-\frac{1}{32}(D_\kappa D_\a F_{\m\n})^{ab}
(D_\l D_\b F_{\r\s})^{cd}\Big\}  \, .\label{NCDejstvo2SSBExp}
\eea
Here $D_\a F_{\m\n}$ is the $SO(2,3)$ covariant derivative and its components
are
\bea
&&(D_\a F_{\m\n})^{ab} = \nabla_\a F_{\m\n}^{\ \ ab} - \frac{1}{l^2}(e_\a
^aT_{\m\n}^{\ b} -e_\a ^bT_{\m\n}^{\ a}) \nn\\
&&(D_\a F_{\m\n})^{a5} = \frac{1}{l}( \nabla_\a T_{\m\n}^{a} + e_\a^m
F_{\m\n m}^{\ \ \ \ a}) \, ,\nn\\
&&(D_\kappa D_\a F_{\m\n})^{ab} = \nabla_\kappa\nabla_\a F_{\m\n}^{\ \ ab} -
\frac{1}{l^2}\Big( 
(\nabla_\kappa e_\a ^a)T_{\m\n}^{\ b} -(\nabla_\kappa e_\a ^b)T_{\m\n}^{\ a} +
e_\a ^a(\nabla_\kappa T_{\m\n}^{\ b})\nn\\
&& - e_\a ^b(\nabla_\kappa T_{\m\n}^{\ a})
+ e_\kappa ^a(\nabla_\a T_{\m\n}^{\ b}) - e_\kappa ^b(\nabla_\a T_{\m\n}^{\ a}) 
+ e_\kappa ^a e^m_\a F_{\m\n m}^{\ \ \ \ b} - e_\kappa ^b e^m_\a F_{\m\n m}^{\ \
\ \ a} \Big) \ ,\nn
\eea
with the $SO(1,3)$ covariant derivative $\nabla_\a F_{\m\n}^{\ \ ab} =
\partial_\a F_{\m\n}^{\ \ ab} + \omega_\alpha^{ac}F_{\mu\nu c}^{\ \ \ b}
-\omega_\alpha^{bc}F_{\mu\nu c}^{\ \ \ a}$ and $\nabla_\a T_{\m\n}^{a} =
\partial_\a T_{\m\n}^{a} + \omega_\alpha^{ac} T_{\mu\nu c}$.

\section{Discussion}

\noindent Starting from the NC $SO(2,3)_\star$ gauge theory, in this paper we
constructed a model of NC gravity. Our construction is based on the enveloping
algebra approach and the SW map. Assuming that the noncommutativity is very
small, we expanded the NC gravity action (\ref{NCdejstvo}) in orders of the
noncommutativity parameter $\theta^{\alpha\beta}$. The first order correction
vanishes; the second order correction is non-zero and we calculated it
explicitly (\ref{NCDejstvo2Exp}). It is obvious that this correction is
invariant under the commutative $SO(2,3)$ gauge transformations. After the
symmetry breaking the action (\ref{NCDejstvo2Exp}) becomes
(\ref{NCDejstvo2SSBExp}) and the symmetry is reduced to the commutative
$SO(1,3)$. The result (\ref{NCDejstvo2SSBExp}) is written in the first order
formalism, the spin connection $\omega_\mu^{ab}$ and the vierbeins $e_\mu^a$ are
independent fields. Unlike in our previous paper, the torsion appears
explicitly in (\ref{NCDejstvo2SSBExp}).

However, the result (\ref{NCDejstvo2SSBExp}) we obtained for the NC gravity
action
is very cumbersome; it is hard to immediately see and discuss any physical
implications from
it. Therefore we analyze different limits of our theory. There are three
different scales in
the model and they are related with the following three parameters: the
cosmological constant $\Lambda=-\frac{3}{l^2}$, the NC parameter
$\theta^{\alpha\beta}$ and
the powers of the curvature tensor (powers of derivatives). Depending on the
values these parameters take different limits of the model are obtained.
Looking closely at (\ref{NCDejstvo2Exp}) we see that separate terms can be
grouped depending on the powers of dimensionless quantity $Rl^2$. There are
five different types:
\begin{equation}
\frac{\theta^2}{l^6}\Big( 1,\ Rl^2,\ R^2l^4,\ R^3l^6,\ R^4l^8\Big) \ .\nn
\end {equation}
At higher energies, higher powers of curvature (which correspond to the
higher powers of derivatives)
are dominant and for some fixed $l$ the leading term is $R^4l^8$. On the other
hand, at low energies, lower powers of curvature are dominant and the
leading term is $R l^2$. The cosmological constant in zeroth order is given by
$\Lambda = -\frac{3}{l^2}$. The limit of the big/small cosmological constant is
obtained taking $l$ to be small/big. The NC parameter $\theta^{\alpha\beta}$ is
taken to be very small, but the dimensionless quantity
$\frac{\theta^2}{l^4}\sim\theta^{2}\Lambda^2$ can be fine-tuned to be
smaller, greater or equal $1$, depending on the value of $l$.
In the following we discuss three different expansions.

\vspace*{0.5cm}
{\bf Expansion 1}
\vspace*{0.3cm}

Let us first assume that we are interested in the limit of small curvature, vanishing
torsion and big cosmological constant. This limit cannot be used to describe
Universe today\footnote{We work with the negative cosmological constant, but all
the results can be easily generalized to the case of positive cosmological
constant.}
but it could have been relevant in some phases of its evolution.
In that case,
from (\ref{NCDejstvo2Exp}) we include only terms which are of zeroth, first and
second order in the curvature. The result is given by
\bea
&&S = S^{(0)} + S^{(2)}\, ,\nn\\
&&S^{(0)} = -\frac{1}{16\pi G_N}\int {\rm d}^4
x\Big[\frac{l^2}{16}\epsilon^{\m\n\r\s}
\epsilon_{abcd}R_{\m\n}^{ab}R_{\r\s}^{cd} + \sqrt{-g}R + 2\sqrt{-g}\Lambda \Big]
\, ,\nn\\
&&S^{(2)} = \frac{3\theta^{\alpha\beta}\theta^{\kappa\lambda}}
{64\pi G_N l^6}\int {\rm d}^4 x \sqrt{-g} g_{\alpha\kappa}g_{\beta\lambda} \nn\\
&&-\frac{\theta^{\alpha\beta}\theta^{\kappa\lambda}}{64\pi G_N l^4}\int {\rm
d}^4 x \sqrt{-g}
\Big( 3g_{\alpha\kappa}R_{\beta\lambda} +3R_{\alpha\beta\kappa\lambda}
-2R_{\alpha\kappa\beta\lambda}  \Big) \nn\\
&& +\frac{\theta^{\alpha\beta}\theta^{\kappa\lambda}}{256\pi G_N l^4}\int {\rm
d}^4 x
\epsilon^{\mu\nu\rho\sigma}\epsilon_{abcd}e_\lambda^{\ c}e_\sigma^{\
d}\nabla_\kappa
\nabla_\alpha (l^2R_{\m\n}^{\ \ ab} -2e_\m^{\ a}e_\n^{\ b})g_{\beta\rho}\nn\\
&& -\frac{\theta^{\alpha\beta}\theta^{\kappa\lambda}}{256\pi G_N l^2}\int {\rm
d}^4 x \sqrt{-g}
\Big( 2g_{\alpha\kappa}( -2RR_{\beta\lambda} +4R_{\beta\mu}R^\mu_{\ \lambda}
+4R^{\mu\nu}R_{\beta\nu\lambda\mu} -2R_{\beta\mu}^{\ \ \
\rho\sigma}R_{\rho\sigma\lambda}^{\
\ \ \ \mu} ) \nn\\
&& + R(2R_{\alpha\kappa\beta\lambda} + R_{\alpha\beta\kappa\lambda})
+2R_{\alpha\kappa}R_{\beta\lambda} -16R_{\alpha\beta\kappa\mu}R^\mu_{\ \lambda}
-16R_{\alpha\kappa\beta\mu}R_\lambda^{\ \mu} \nn\\
&& -2R_{\alpha\beta}^{\ \ \ \mu\nu}( R_{\kappa\lambda\mu\nu} -
6R_{\kappa\mu\lambda\nu})
+2R_{\alpha\kappa}^{\ \ \ \mu\nu}( 5R_{\beta\lambda\mu\nu}
-4R_{\beta\mu\lambda\nu})\nn\\
&&+4 R_{\alpha\mu\beta}^{\ \ \ \ \nu}R_{\kappa\nu\lambda}^{\ \ \ \ \mu} 
+ 4 R_{\alpha\mu\kappa}^{\ \ \ \ \nu}R_{\lambda\nu\beta}^{\ \ \ \ \mu}
-6R_{\alpha\mu\kappa}^{\ \ \ \ \nu}R_{\beta\nu\lambda}^{\ \ \ \ \mu}
\Big) \nn\\
&& +\frac{\theta^{\alpha\beta}\theta^{\kappa\lambda}}{256\pi G_N l^2}\int {\rm
d}^4 x
\epsilon^{\mu\nu\rho\sigma}\epsilon_{abcd}\Big\{ -2e_\a^{\ a}e_\b^{\
b}(\nabla_\kappa
(R_{\m\n}^{\ \ cm})\nabla_\l( e_{\r m}e_\s^{\ d} ) \nn\\
&& + \frac{2}{l^2}\nabla_\k(e_\m^{\ c}e_\n^{\ m})\nabla_\lambda(e_{\r m}e_\s^{\
d})) \label{S2}\\
&& +e_\r^{\ a}e_\s^{\ b}\Big[ 2\nabla_\kappa R_{\a\m}^{\ \ cm}\nabla_\l (e_{\b
m}e_\n^{\ d} - e_\b^{\ d}e_{\n m})\nn\\
&&- \frac{1}{l^2}\nabla_\k(e_\a^{\ c}e_\m^{\ m} - e_\a^{\ m}e_\m^{\ c})
\nabla_\lambda(e_{\b m}e_\n^{\ d} - e_\b^{\ d}e_{\n m})\Big]\nn\\
&& +\frac{1}{2} \nabla_\kappa \nabla_\alpha (e_\m^{\ a}e_\n^{\ b})
\big(\nabla_\lambda \nabla_\beta (e_\r^{\ c}e_\s^{\ d}) 
+ 2e_\lambda^{\ c}R_{\r\s\b}^{\ \ \ \ \omega}e_\omega^{\ d}\big) \Big\}   \nn\\
&& -\frac{\theta^{\alpha\beta}\theta^{\kappa\lambda}}{256\pi G_N l^2}\int {\rm
d}^4 x
\epsilon^{\mu\nu\rho\sigma}\epsilon_{abcd}\Big\{ R_{\a\b}^{\ \
ab}\nabla_\k(e_\m^{\ c}e_\n^{\ m})
\nabla_\lambda(e_{\r m}e_\s^{\ d}) \nn\\
&&-\frac{1}{2}R_{\r\s}^{\ \ ab}\nabla_\k(e_\a^{\ c}e_\m^{\ m} - e_\a^{\
m}e_\m^{\ c})
\nabla_\lambda(e_{\b m}e_\n^{\ d} - e_\b^{\ d}e_{\n m}) \Big\}
 \, .\nn
\eea
This action is invariant under the $SO(1,3)$ gauge symmetry. However, due to the
noncommutativity it is no
longer invariant under the diffeomorphism symmetry. The non-invariant terms
manifest in two ways. Firstly, there are tensors contracted with the NC
parameter $\theta^{\a\b}$ such as
$\theta^{\a\b}\theta^{\k\l}R_{\alpha\kappa\beta\lambda}$. Since $\theta^{\a\b}$
is not a tensor under the diffeomorphism symmetry (it is a non-transforming
constant matrix), those terms are also not tensors. Then there are terms in
which $SO(1,3)$ covariant derivatives of vierbeins appear. Using the metricity
condition 
\begin{equation}
\nabla_\mu^{tot} e_\rho^{\ a} = \partial_\mu e_\rho^{\ a} + \omega_\mu^{ab} e_{\rho b} - \Gamma_{\mu\rho}^\sigma e_\sigma^{\ a} =0 
\end{equation}
the $SO(1,3)$ covariant derivative can be written as
\begin{equation}
\nabla_\mu e_\rho^{\ a} = \partial_\mu e_\rho^{\ a} + \omega_\mu^{ab} e_{\rho b} = \Gamma_{\mu\rho}^\sigma e_\sigma^{\ a} \ . \label{LorKovIzvTetrada}
\end{equation}
Therefore the affine connection (Christoffel symbols) $\Gamma_{\mu\rho}^\sigma$
appears explicitly in (\ref{S2}).
Some of the terms can be grouped to to the curvature tensor, but some will
remain and make the diffeomorphism non-invariance explicit. On the other hand,
the action (\ref{NCdejstvo}) is invariant under the twisted diffeomorphisms. How
the symmetry breaking affects this invariance is not clear yet and remains to be
studied further.

The assumption of the big cosmological constant leads to $1\gg l^2R\gg
\frac{\theta^2}{l^4}$ and therefore selects the leading order
terms to be proportional to $1/l^6$. Explicitly
\begin{equation}
S^{(2)} = \frac{3\theta^{\alpha\beta}\theta^{\kappa\lambda}}
{64\pi G_N l^6}\int {\rm d}^4 x \sqrt{-g} g_{\alpha\kappa}g_{\beta\lambda} \ .
\label{S2model1}
\end{equation}
The equation of motion following from (\ref{S2model1}) is obtained varying with
respect to $g_{\mu\nu}$ and it is given by
\begin{equation}
R_{\r\s} -\frac{1}{2}g_{\r\s}(R+2\Lambda)
+\frac{3}{4l^6}\theta^{\a\b}\theta^{\k\l}(\frac{1}{2}g_{\r\s}g_{\a\k}g_{\b\l} +
2g_{\b\l}g_{\a\r}g_{\k\s})  =0 \ . \label{EOMmodel1} 
\end{equation}
The cosmological constant is modified in this model, it becomes $x$-dependent
\begin{equation}
\Lambda (x) = \Lambda
-\frac{3}{8}\frac{\theta^{\a\b}\theta^{\k\l}}{l^6}g_{\a\k}g_{\b\l} \ .
\label{Lambda(x)} 
\end{equation}
To see if this modification can "flatten" the starting commutative space (AdS is
the zeroth order solution of equations (\ref{EOMmodel1})) we check whether the
Minkowski space-time is a solution of (\ref{EOMmodel1}). That is, we look for
$\theta^{\a\b}$ such that $g_{\mu\nu}=\eta_{\mu\nu}$ is a solution of
(\ref{EOMmodel1}). Unfortunately, this is not the case. The easiest way
to see this is to look at the equation obtained by contracting (\ref{EOMmodel1})
\begin{equation}
R = 4\Lambda
+\frac{3}{l^6}\theta^{\alpha\beta}\theta^{\kappa\lambda}g_{\alpha\kappa}g_{
\beta\lambda} \ . \label{EOMmodel1Contracted} 
\end{equation}
Demanding that $g_{\mu\nu}=\eta_{\mu\nu}$ is a solution of this equation, gives
$\frac{\theta^2}{l^4}=2$, which is in contradiction with the assumption that
$1\gg l^2R\gg \frac{\theta^2}{l^4}$. Therefore we conclude that the Minkowski
space-time cannot be a solution of (\ref{EOMmodel1}). 

We know that the zeroth order solution of (\ref{EOMmodel1}) is AdS space-time,  $g_{\m\n}^{(0)}
= g_{\m\n}^{AdS}$.  Making
an expansion $g_{\mu\nu}=g_{\mu\nu}^{AdS} + \varepsilon h_{\mu\nu}$ with a small
parameter $\varepsilon$ we can linearize the equations around this solution. This has to be done very carefully and we postpone the
calculation for the next paper. 

Note that in our previous paper \cite{MiAdSGrav} we did not obtain a $x$-dependent correction to the cosmological constant. This shows once more that the deformation and symmetry breaking do not commute, instead they lead to different models.

\vspace*{0.5cm}
{\bf Expansion 2}
\vspace*{0.3cm}

Our next choice is the limit of the small cosmological constant and vanishing
torsion. It can be relevant when describing Universe today or in some of its
earlier phases. In that case the second order correction for the NC gravity
action is given by
\bea
S^{(2)}&=&-\frac{l^2\theta^{\alpha\beta}\theta^{\kappa\lambda}}{64\pi G_N}\int
d^4
xe\bigg [\nn\\
&&\Big(-\frac{1}{64}R_{\a\b\g\d}R_{\kappa\l}^{\ \
\g\d}+\frac{1}{32}R_{\kappa\a\g\delta}R_{\l\b}^{\ \
\g\delta}\Big)\Big(R^2+4R_{\m\n}R^{\m\n}+R_{\m\n\r\s}R^{\m\n\r\s}\Big)\nn\\
&& +R_{\m\n\g\d}R^{\r\s\g\d}\Big(\frac{1}{32}(2
R_{\a\b\ \r}^{\ \ \m}R_{\k\l\ s}^{\ \ \n}-
R_{\a\b}^{\ \ \m\n}R_{\k\l\r\s})\nn\\
&&+\frac{1}{16}R_{\kappa\a\r\s}
R_{\l\b}^{\ \ \m\n}-\frac18R_{\k\a\n\s}R_{\l\b\ \r}^{\ \ \m}\Big)\nn\\
&&+\frac14R_{\k\l\g\d}R^{\r\s\g\d}(R_{\a\m\r\s}R_\b^{\ \m}+R_{\a\m\r\n}R_{\b\
\s}^{\ \n\ \m}-R_{\a\r}R_{\b\s})\nn\\
&&+\frac18R_{\a\m\g\d}R_{\b\n}^{\ \ \g\d}(R
R_{\k\l}^{\ \ \m\n}+R_{\k\l\r\s}R^{\r\s\m\n}+4R_{\k\l\ \r}^{\ \
\n\ }R^{\r\m})\nn\\
&&+\frac12R_{\a\r\g\d}R_{\b\s}^{\ \ \g\d}(R_{\k\m}^{\ \ \r\s}R_\l^{\
\m}+R_{\k\m\n\r}R_\l^{\ \n\m\s}+R_{\k\s}R_{\l\r})\nn\\
&&-\frac14R_{\k\a\g\d}
R_{\l\m}^{\ \ \g\d}(R_{\b\n\r\s}R^{\r\s\m\n}+RR_{\b\m}-2R_{ \s\m}
R_\b^{\ \s}-2R_{\s\n}R_\b^{\ \n\m\s} )\nn\\
&&-\frac{1}{4}R_{\r\s\g\d}R_{\b\n}^{\g\d}(R_{\k\a}^{\ \ \m\n}R_{\l\m}^{\
\ \r\s}+2R_{\k\a}^{\ \ \m\s}R_{\l\m}^{\ \ \n\r}+2R_{\k\a}^{\ \
\r\n}R_\l^{\ \s}-R_{\k\a}^{\ \ \r\s}R_{\l\n}
)\bigg] \nn\\
&&-\frac{l^2\theta^{\alpha\beta}\theta^{\kappa\lambda}}{64\pi
G_N}\epsilon^{\m\n\r\s}\epsilon_{abcd}\int
d^4x\bigg[\frac{1}{16}R_{\a\b}^{ab}
(\nabla_\k R_{\m\n})^{cm}(\nabla_\l R_{\r\s})_m^{\
d}\nn\\
&&-\frac{1}{8}R_{\r\s}^{ab}
(\nabla_\k R_{\a\m})^{cm}(\nabla_\l R_{\b\n})_m^{\ d}
-\frac{1}{32}
(\nabla_\k \nabla_\a R_{\m\n})^{ab}(\nabla_\l \nabla_\b R_{\r\s})^{cd}
\Big)\bigg] \ . \label{S2model2}
\eea
The correction terms are of the third and fourth power of curvature. Since the
cosmological constant is small, we can assume that the zeroth order solution is
the Minkowski space-time, $g^{(0)}_{\mu\nu}\approx\eta_{\mu\nu}$. Then we can expand
around this solution assuming $g_{\mu\nu} = \eta_{\mu\nu} + \varepsilon h_{\mu\nu}$ and
look for the equation for $h_{\mu\nu}$ and its solutions. This we postpone for
the next paper.

\vspace*{0.5cm}
{\bf Expansion 3}
\vspace*{0.3cm}

Our final example is the "NC teleparallel" solution: we
assume $R_{\m\n}^{\ \ ab}=0$ and $T_{\m\n}^a\ne0$. The second order correction
is given by
\begin{eqnarray}
S_T&=&
 -\frac{\theta^{\alpha\beta}\theta^{\kappa\lambda}}
{64\pi G_N l^6}\int {\rm d}^4 x \sqrt{-g}\Big[
5T_{\kappa\a}^mT_{\l\b m}
-\frac{11}{4}T_{\a\b}^m T_{\k\l m}-\frac{3}{2}T_{\r\kappa}^\r T_{\a\b
\l}\nn\\
&&+T_{\a\l}^\n T_{\b\n
\kappa}-5T_{\a\kappa\b}T_{\m\l}^\m+2T_{\a\b}^\m T_{\kappa\m\l }
+\frac12g_{\l\n} e^\n_c\pa_\kappa T_{\a\b}^c-g_{\l\a} e^\n_a\pa_\kappa
T_{\n\b}^a\nn\\&-&\frac12(e_c^\m e_d^\n-e_c^\n e_d^\m)\pa_\kappa(e_\a^c
e_\m^m-e_\a^m e_\m^c)\pa_\l(e_{\b m} e_\n^d-e_\b^d e_{\n m})\nn\\
&&-\pa_\kappa(e_\a^c e_\m^m-e_\a^m
e_\m^c)\Big( e_c^\m(e_{\l m}T_{\m\n}^\n-T_{\b\l m})-e_c^\n(e_{\l
m}T_{\b\n}^\m-\d_\l^\m T_{\b\n m})\Big) \nn\\
&&+\frac12l^2(e_c^\m e_d^\n-e_d^\m e_c^\n)\pa_\kappa T_{\a\m}^c\pa_\l
T_{\b\n}^d\Big]\nn\\
&& -\frac{\theta^{\alpha\beta}\theta^{\kappa\lambda}}
{64\pi G_N l^6}\epsilon^{\mu\nu\rho\sigma}\epsilon_{abcd}\int {\rm d}^4 x
\Big[\nn\\
&&-\frac12  e_\a^a e_\b^b\pa_\k(e_\m ^c e_{\n m}) \Big( \pa_\l(e_{\r
m}e_\s^d)+e_{\l
m}T_{\r\s}^d-e_\l^d T_{\r\s m}\Big) +\frac12\pa_\a\pa_\kappa(e_\m^a e_\n^b)e_\l
^c e_\s^d g_{\b\r}\nn\\
&& +2\pa_\a(e_\l^b e_\n^c)g_{\b\r}e_\s^d T_{\kappa\m}^a+\frac12 \pa_\l(e_\m^b
e_\n^c)g_{\b\r}e_\s^d T_{\kappa\a}^a +2e_\k ^a e_\m ^b e_\s^d g_{\b\r}\pa_\a
T_{\l\n}^c\nn\\
&& +e_\kappa^a e_\m^b g_{\a\l}e_\n ^c\pa_\b T_{\r\s}^d+\frac12
e_\kappa ^a e_\a^b (e_\s^d g_{\b\r}\pa_\l T_{\m\n}^c+e_\n^c g_{\l\m}\pa_\b
T_{\r\s}^d)+ e_\kappa ^a e_\n^b e_\l^c g_{\a\m}\pa_\b T_{\r\s}^d\nn\\
&& -\frac14\pa_\a\pa_\k(e_\m^a e_\n^b)e_\l^c \pa_\b
T_{\r\s}^d-\frac14\pa_\k(e_\a^a T_{\m\n}^b)e_\l^c \pa_\b T_{\r\s}^d\nn\\
&& +\frac12 e_\l^c e_\s^dg_{\b\r}\pa_\kappa(e_\a^a  T_{\m\n}^b)-\frac14e_\kappa
^ae_\l^c\pa_\a T_{\m\n}^b\pa_\b T_{\r\s}^d
-\frac14g_{\a\l}e_\b^d\pa_\kappa(e_\m^a
e_\n^b)T_{\r\s}^c\nn\\
&&-\frac14e_\kappa^ae_\a^b\pa_\l T_{\m\n}^c\pa_\b T_{\r\s}^d
+\frac14T_{\kappa\a}^a(-\pa_\l(e_\m^b e_\n^c)\pa_\b
T_{\r\s}^d+e_\l^cT_{\m\n}^b\pa_\b T_{\r\s}^d)
\Big] \ . \label{S2model3}
\end{eqnarray}
The zeroth order action in this limit reduces to the cosmological constant
term. That means that there is no kinetic term for the vierbeins in the zeroth
order. We would have to look for it in (\ref{S2model3}). Another way to obtain
it would be to start form a different commutative action, the one that includes
the kinetic term for vierbeins. 

The analysis of diffeomorphism invariance of (\ref{S2model2}) and
(\ref{S2model3}) remains the same as in the first example (\ref{S2}): the
commutative diffeomorphism symmetry is broken, while the invariance under the
twisted diffeomorphism symmetry remains to be understood better.

Finally, let us comment that our results (\ref{S2}), (\ref{S2model2}) and
(\ref{S2model3}) cannot be related with $f(R)$ and $f(T)$ theories. Some of the
indices on the curvature tensor and torsion will always be contracted with the
NC parameter $\theta^{\a\b}$, this is a consequence of the SW map. Therefore, it
seems impossible to construct invariants of curvature tensor or torsion alone.

\vskip1cm \noindent
{\bf Acknowledgement}
\hskip0.3cm
We would like to thank Milutin Blagojevi\' c, Maja Buri\' c and Hrvoje \v
Stefan\v ci\' c for
fruitful discussion and useful comments. The work is
supported by project
171031 of the Serbian Ministry of Education and Science and partially by
ICTP-SEENET-MTP
Project PRJ09 ”Cosmology and Strings” in frame of the Southeastern European
Network in Theoretical and Mathematical Physics.

\appendix

\section{AdS algebra and the $\gamma$-matrices}

Algebra relations\footnote{$\epsilon^{01235}=+1,\ \epsilon^{0123}=1$}:
\bea \{M_{AB},\Gamma_C\}&=&i\epsilon_{ABCDE}M^{DE}\nn\\
\{M_{AB},M_{CD}\}&=&\frac{i}{2}\epsilon_{ABCDE}\Gamma^{E}+\frac12(\eta_{AC}\eta_
{BD}-\eta_{AD}\eta_{BC})\nn\\
{[}M_{AB},\Gamma_C{]}&=&i(\eta_{BC}\Gamma_A-\eta_{AC}\Gamma_B)\nn\\
\Gamma_A^\dagger&=&-\gamma_0\Gamma_A\gamma_0\nn\\
M_{AB}^\dagger&=&\gamma_0M_{AB}\gamma_0\nn\\
\{\sigma_{ab},\sigma_{cd}\}&=&2(\eta_{ac}\eta_{bd}-\eta_{ad}\eta_{bc}+i\epsilon_
{abcd}\gamma_5)\nn\\
{[}\sigma_{ab},\gamma_c{]}&=&2i(\eta_{bc}\gamma_a-\eta_{ac}\gamma_b)\nn\\
\{\sigma_{ab},\gamma_c\}&=&2\epsilon_{abcd}\g^5\g^d
\eea
Identities with traces:
\bea
&&\tr (\Gamma_A\Gamma_B)=4\eta_{AB}\nn\\
&&\tr (\Gamma_A)=\tr (\Gamma_A\Gamma_B\Gamma_C)=0\nn\\
&&\tr
(\Gamma_A\Gamma_B\Gamma_C\Gamma_D)=4(\eta_{AB}\eta_{CD}-\eta_{AC}\eta_{BD}+\eta_
{AD}\eta_{CB})\nn\\
&&\tr (\Gamma_A\Gamma_B\Gamma_C\Gamma_D\Gamma_E)=-4i\epsilon_{ABCDE}\nn\\
&&\tr (M_{AB}M_{CD}\Gamma_E)=i\epsilon_{ABCDE}\nn\\
&&\tr (M_{AB}M_{CD})=-\eta_{AD}\eta_{CB}+\eta_{AC}\eta_{BD}
\eea


\begin{thebibliography}{99}

\bibitem{Doplicher} S. Doplicher, K. Fredenhagen and J. E. Roberts, {\it The
Quantum structure of space-time at the Planck scale and quantum fields},
Commun.\ Math.\ Phys. {\bf 172}, 187 (1995), [hep-th/0303037].

\bibitem{NCbooks}
A.~Connes, {\it Non-commutative Geometry}, Academic Press (1994).

J. Madore, {\it An Introduction to Noncommutative Differential
Geometry and its Physical Applications}, 2nd Edition, Cambridge
Univ. Press (1999).

P.~Aschieri, M.~Dimitrijevi\' c, P.~Kulish, F.~Lizzi and J.~Wess
{\it Noncommutative spacetimes:
Symmetries in noncommutative geometry and field theory},
Lecture notes in physics {\bf 774}, Springer (2009).

\bibitem{defgt}
P.~Aschieri, C.~Blohmann, M.~Dimitrijevi\' c, F.~Meyer, P.~Schupp
and J.~Wess, {\it A Gravity Theory on Noncommutative Spaces},
Class.\ Quant.\ Grav. {\bf 22}, 3511 (2005), [hep-th/0504183 ].

P.~Aschieri, M. Dimitrijevi\' c, F.~Meyer and  J.~Wess,
{\it Noncommutative Geometry and Gravity}, Class.\ Quant.\ Grav. {\bf
23}, 1883 (2006), [hep-th/0510059].

\bibitem{Cha01} A. H. Chamseddine, {\it Complexified gravity in noncommutative
spaces}, Commun.\ Math.\ Phys. {\bf 218}, 283 (2001) [hep-th/0005222].

\bibitem{ChPL} A. H. Chamseeddine, {\it Deforming Einstein's gravity},
Phys.\ Lett.\ B {\bf 504} 33 (2001), [hep-th/0009153].

\bibitem{Cha04} M. A. Cardella and D. Zanon,
{\it Noncommutative deformation of four-dimensional gravity}, Class.\ Quant.\
Grav. {\bf 20}, L95 (2003), [hep-th/0212071].

A. H. Chamseddine, {\it $SL(2,c)$ gravity with complex vierbein and its
noncommutative extension}, Phys.\ Rev.\ {\bf D69}, 024015
(2004), [hep-th/0309166].

\bibitem{PC09} P. Aschieri and L. Castellani, {\it Noncommutative $D=4$ gravity
coupled to fermions} JHEP, {\bf 0906}, 086 (2009), [arXiv:0902.3823].

\bibitem{Ste10} 
H. S. Yang, {\it Emergent gravity from noncommutative spacetime}, Int.\ J.\
Mod.\ Phys. {\bf A24}, 4473 (2009), [hep-th/0611174].

H. Steinacker, {\it Emergent Geometry and Gravity from Matrix Models: an
Introduction}, Class.\ Quant.\ Grav.\ {\bf 27}, 133001
(2010), [arXiv:1003.4134].

\bibitem{MadoreMaja}
M. Buri\' c and J. Madore, {\it Spherically Symmetric Noncommutative Space: $d =
4$}, Eur.\ Phys.\ J. {\bf C58},  347 (2008), [arXiv: 0807.0960].

M. Buri\' c and J. Madore, {\it On noncommutative spherically symmetric spaces},
arXiv:1401.3652.


\bibitem{ChamMukh}
A. H. Chamseddine and V. Mukhanov, {\it Who Ordered the Anti-de Sitter Tangent
Group?},  JHEP {\bf 1311} 095 (2013), [arXiv:1308.3199].

A. H. Chamseddine and V. Mukhanov, {\it Gravity with de Sitter and Unitary
Tangent Groups}, JHEP {\bf 1003}, 033 (2010), [arXiv:1002.0541].

\bibitem{Barrett}
J. W. Barrett and S. Kerr, {\it Gauge gravity and discrete quantum models},
arXiv:1309.1660.

\bibitem{AdSCFT}
O. Aharony, S.S. Gubser, J. Maldacena, H. Ooguri and Y. Oz, {\it Large N Field
Theories, String Theory and Gravity}, Phys.\ Rept.\ {\bf 323}, 183 (2000),
[hep-th/9905111].

\bibitem{MiAdSGrav}
M. Dimitrijevi\' c, V. Radovanovi\' c and H. \v Stefan\v ci\' c,
{\it AdS-inspired noncommutative gravity on the Moyal plane},
Phys. Rev. D {\bf 86}, 105041 (2012), [arXiv:1207.4675].

\bibitem{jssw}
B. Jur\v{c}o, S.~Schraml, P.~Schupp and J.~Wess,
{\it Enveloping algebra valued gauge transformations for
non-abelian gauge groups on non-commutative spaces }, Eur.\ Phys.\ J.\ C{\bf
17}, 521 (2000), [hep-th/0006246].

B. Jur\v{c}o, L. M\"oller, S.~Schraml, P.~Schupp and J.~Wess,
{\it Construction of non-Abelian gauge theories on noncommutative spaces},
Eur.\ Phys.\ J.\ C{\bf 21}, 383 (2001), [hep-th/0104153].

\bibitem{Seiberg:1999vs}
N.~Seiberg and E.~Witten,
{\it String theory and noncommutative geometry},
JHEP {\bf 09}, 032 (1999), [hep-th/9908142].

\bibitem{PLM} P. Aschieri, L. Castellani and M. Dimitrijevi\'c, {\it
Noncommutative gravity at second order
via Seiberg-Witten map}, Phys. Rev. D {\bf 87},  024017 (2013),
[arXiv:1207.4346]

\bibitem{stelle-west} K. S. Stelle and P. C. West, {\it Spontaneously
broken de Sitter symmetry and the gravitational holonomy group}, Phys.\ Rev D
{\bf 21}, 1466 (1980).

\bibitem{McD-Mansouri} S. W. MacDowell and F. Mansouri,
{\it Unified geometrical theory of gravity and supergravity}, Phys.\ Rev.\ Lett.
{\bf 38}, 739 (1977).

\bibitem{Towsend} P. K. Towsend, {\it Small-scale structure of spacetime as
the origin of the gravitation constant}, Phys.\ Rev.\ D {\bf 15},  2795 (1977).

\bibitem{GHY} J. W. York, {\it Role of conformal three-geometry in the
dynamics of gravitation}, Phys.\ Rev.\ Lett. 28, (1972). 

G. W. Gibbons, S. W. Hawking, {\it Action integrals and partition functions
in quantum gravity}, Phys.\ Rev. {\bf D 15}, 2752 (1977).

\bibitem{Wald} R. Wald, {\it General Relativity}, University of Chicago Press,
Chicago and London (1984).

\bibitem{kayahan} K. Ulker and B. Yapiskan, {\it Seiberg-Witten maps to all
orders}, Phys.\ Rev.\ D {\bf 77}, 065006 (2008), [arXiv: 0712.0506].

\bibitem{PC11} P. Aschieri and L. Castellani, {\it Noncommutative gravity 
coupled to fermions: second order expansion via Seiberg-Witten map},  JHEP {\bf
1207} 184 (2012), [arXiv:1111.4822].

\bibitem{Peskin} T. P. Sotiriou, S. Liberati, {\it Field
equations from a surface term}, Phys.Rev. {\bf D74}, 044016 (2006). 

\bibitem{MilutinKnjiga} M. Blagojevi\'c, {\it Gravitation and Gauge Symmetries},
Institute of Physics Publication, Bristol (2002).



\bibitem{Mukherjee2} P. Mukherjee and A. Saha, {\it A Note on the noncommutative
correction to gravity},
Phys.\ Rev D {\bf 74}, 027702 (2006), [hep-th/0605287].




\end{thebibliography}
\end{document}